\documentclass[fleqn,twoside,11pt]{article}

\usepackage{amssymb}
\usepackage{graphicx}
\usepackage{amsmath}

\begin{document}

\begin{center}

{\Large\bf Neutrino Mixing based on Mass Matrices }

\vspace{2mm}
{\Large\bf with a $2 \leftrightarrow 3$ Symmetry}

\vspace{3mm}
{\bf Yoshio Koide and Eiichi Takasugi}

{\it Institute for Higher Education Research and Practice, Osaka University, \\
1-16 Machikaneyama, Toyonaka, Osaka 560-0043, Japan\\

}

\date{\today}
\end{center}

\begin{abstract}
Under the assumption that the $2\leftrightarrow 3$ symmetry is broken
only through phases, we give a systematical investigation  of possible
lepton mass matrix forms without referring to the explicit parameter values.
The two types of the $2\leftrightarrow 3$ symmetry are investigated:
one is that the left- and right-handed fields $(f_L, f_R)$ obey the 
symmetry, and another one is that  only $f_L$ obeys the symmetry.
In latter case, in spite of no $2\leftrightarrow 3$ symmetry in the
Majorana mass matrix $M_R$ for $\nu_R$, the neutrino seesaw mass
matrix still obey the $2\leftrightarrow 3$ symmetry.   Possible
phenomenologies are discussed.
\end{abstract}

\vspace{3mm}
\noindent{\large\bf 1 \ Introduction}

We usually consider that the quarks and leptons should be understood by a
unification theory.
Then, the concept of ``symmetry" will become important 
in the understanding of ``flavor".
It is well-known that the requirement of the $2\leftrightarrow 3$ 
symmetry \cite{23sym} 
for the neutrino mass matrix leads to the maximal mixing between the $\nu_2$
and $\nu_3$ components.
The idea of the the $2\leftrightarrow 3$ symmetry
is very promising for understanding the observed neutrino mixing.

When a matrix $M$ satisfies the relation
$$
T_{23} M T_{23}^\dagger = M ,
\eqno(1.1)
$$
where $T_{23}$ is defined as
$$
T_{23} = \left( 
\begin{array}{ccc}
1 & 0 & 0 \\
0 & 0 & 1 \\
0 & 1 & 0
\end{array} \right) ,
\eqno(1.2)
$$
the matrix $M$ is called ``$2\leftrightarrow 3$ symmetric".
The mass matrix form of $M$ is explicitly expressed as
$$
M=\left(
\begin{array}{ccc}
d & a & a \\
a & b & c \\
a & c & b
\end{array} \right) .
\eqno(1.3)
$$

Firstly, we would 
like to notice that the mass matrix which satisfies Eq.(1.1)  
is considered a consequence of the invariance of 
the mass matrix under the field transformation. Explicitly, 
for  the Dirac mass matrix $\bar{f}_L M f_R$,  
Eq.(1.1) is derived by requiring 
the invariance under  
the transformation, $f_L \rightarrow T_{23}^\dagger f_L$ and 
$f_R \rightarrow T_{23}^\dagger f_R$. 
This is true if the neutrino mass matrix is 
derived by the seesaw mechanism, $M_\nu=m_DM_R^{-1}m_D^T$ because this 
matrix $M_\nu$ is invariant under $f_R \rightarrow Tf_R$ with any $T$. 
Next, we 
focus on the transformation of $\nu_L$ and $e_L$. Since they 
forms an doublet of the electroweak symmetry, the transformation for 
them should be the same. That makes a big trouble to realize the 
reasonable neutrino mixing as we see in the next section. 

Now we extend the $2 \leftrightarrow 3$ symmetry according to 
multiplets  under the electroweak symmetry. In general, 
the transformation between $(\nu_L,e_L)$ and $\nu_R$ are 
different. This is true even we consider the SU(5) GUT. 
On the other hand, in the SO(10) GUT, $(\nu_L,e_L)$ and 
$(\nu_R, e_R)$ 
will be transformed under the same operator $T_{23}$.
According to this classification, two types of $2 \leftrightarrow 3$ 
symmetry arises. 
The one 
(we call it Type I) is that both $f_L$ and $f_R$ obey the
$2 \leftrightarrow 3$ symmetry.  
Eq.(1.3) is obtained for charged leptons and also for neutrinos.  
Consider that $T_{23}M_L T_{23}^\dagger = M_L$ and 
$T_{23}M_R T_{23} = M_R$ and we find the Majorana mass matrix in 
Eq.(1.3). 

The other one (we call it Type II) is the case where only $f_L$ obeys
the $2 \leftrightarrow 3$ symmetry. Then, we find for the Dirac 
mass matrix $M_L^f$ (we define a Dirac mass matrix $M_L^f$ as
$\bar{f}_L M_L^f f_R$) 
$$
T_{23} M_L^f  = M_L^f .
\eqno(1.4)
$$
and the explicit form of the mass matrix $M_L^f$ is given by
$$
M_L^f = \left(
\begin{array}{ccc}
a_1 & b_1 & c_1  \\
a   &  b  &  c  \\
a   &  b  &  c 
\end{array} \right) .
\eqno(1.5)
$$
The neutrino mass matrix $M_\nu = M_L^\nu M_R^{-1} (M_L^\nu)^T$
is given as a special case of Eq.(1.3) 
by taking $b=c$ as we shall see later.

Note that, in the both types I and II, the Hermitian matrix defined by 
$H_f =M_f M_f^\dagger$ satisfies the constraint
$$
T_{23} H_f T_{23}^\dagger =H_f ,
\eqno(1.6)
$$
independently whether the mass matrx has the form in Eq.(1.3) or (1.5). 

Now the neutrino mixing matrix $U$ is given by
$$
U =U_{Le}^\dagger U_{L\nu} ,
\eqno(1.7)
$$
where $U_{Lf}$ are defined by
$$
U_{Lf}^\dagger H_f U_{Lf} = {\rm diag}(m^2_{f1}, m^2_{f2}, 
m^2_{f3}) \equiv D^2_f .
\eqno(1.8)
$$
From the argument given above, we learned that 
as far as the mixing matrix $U$ is concerned, the structure of the 
neutrino mixing 
matrix is independent of the mass matrices of Types I or II. 
Only difference arises in the mass spectrum. 
 
The purposes of the present paper is to investigate the general
properties of the models with the $2 \leftrightarrow 3$ symmetry, 
paying attention to the difference between types I and II, 
and taking relations to the grand unification (GUT) scenarios 
into consideration.
Although we investigate the masses and mixings
in the lepton sectors, the formulation in this paper is also
applicable to the quark sectors. 
Since, in the quark sectors, there is essentially no case 
complexity about the mass spectrum such as the inverse 
hierarchy as in the neutrino sector, the application is more
straightforward. Therefore, 
we will investigate only the lepton sectors in this paper.

\vspace{3mm}
\noindent{\large\bf 2 \  Extended  $2\leftrightarrow 3$ symmetry and the 
neutrino mixing}

In this section, we will demonstrate that  the $2\leftrightarrow 3$
symmetry in the exact meaning cannot explain the observed
neutrino mixing. 
For the convenience of the discussion in later, let us introduce the
so-called extended $2\leftrightarrow 3$ operator $T_{23}(2\delta)$
\cite{ext23sym}
$$
T_{23}(2\delta) = \left(
\begin{array}{ccc}
1 & 0 & 0 \\
0 & 0 & e^{i2\delta} \\
0 & e^{-i2\delta} & 0
\end{array} \right) ,
\eqno(2.1)
$$
instead of the operator (1.2) and consider both types. 
The operator $T_{23}(2\delta)$ is unitary and Hermitian. 
We obtain the constraint
$$
T_{23}(2\delta) M M^\dagger T_{23}^\dagger(2\delta)
= M M^\dagger ,
\eqno(2.2)
$$
for the Hermitian matrix $M M^\dagger$
irrespective of Type I or II.
Note that we can express the operator (2.1) as
$$
T_{23}(2\delta) = P_{23}(2\delta) T_{23} 
= P_{23}(\delta) T_{23}P_{23}^\dagger(\delta)
= T_{23}P_{23}^\dagger(2\delta) ,
\eqno(2.3)
$$
where $T_{23} =T_{23}(0)$ and
$$
P_{23}(\delta) = {\rm diag} (1, e^{i\delta}, e^{-i\delta}).
\eqno(2.4)
$$
Therefore, we can express  the constraint (2.2) as
$$
M M^\dagger = P_{23}(\delta) T_{23}P_{23}^\dagger(\delta)
M M^\dagger P_{23}(\delta) T_{23}P_{23}^\dagger(\delta).
\eqno(2.5)
$$
Now we define
$$
H= P_{23}^\dagger(\delta)
M M^\dagger P_{23}(\delta),
\eqno(2.6)
$$
then we find
$$
H = T_{23}H T_{23} , 
\eqno(2.7)
$$
where $H$ is a Hermitian matrix

In general, the Hermitian matrix $H$ which satisfies the 
constraint (2.6) can be expressed by the form
$$
H= \left( 
\begin{array}{ccc}
D & A e^{i \phi} & A e^{i \phi} \\
A e^{-i \phi} & B & C \\
A e^{-i \phi} & C & B 
\end{array} \right) ,
\eqno(2.8)
$$
where $A$, $B$, $C$ and $D$ are real, 
so that $H$ can be transformed to a real matrix $\widetilde{H}$
as
$$
P_1^\dagger(\phi) H P_1(\phi) = \widetilde{H} ,
\eqno(2.9)
$$
where
$$
P_1(\phi) = {\rm diag}(e^{i\phi}, 1, 1) .
\eqno(2.10)
$$
It is also well-known that the $2\leftrightarrow 3$
symmetric real matrix $\widetilde{H}$ is diagonalized by a
rotation $R(\theta)$ as
$$
R^T(\theta) \widetilde{H} R(\theta) = \widetilde{H}_D
\equiv {\rm diag}(m_1^2, m_2^2, m_3^2) ,
\eqno(2.10)
$$
where
$$
R(\theta) = \left(
\begin{array}{ccc}
c_\theta & s_\theta & 0 \\
-\frac{1}{\sqrt2} s_\theta &  \frac{1}{\sqrt2} c_\theta
& -\frac{1}{\sqrt2} \\
-\frac{1}{\sqrt2} s_\theta &  \frac{1}{\sqrt2} c_\theta
& \frac{1}{\sqrt2}
\end{array} \right) ,
\eqno(2.11)
$$
$$
s_\theta \equiv \sin\theta =
\sqrt{\frac{D-m_1^2}{m_2^2 -m_1^2}} , \ \ \ 
c_\theta \equiv \cos\theta =
\sqrt{\frac{m_2^2 - D}{m_2^2 -m_1^2}} , 
\eqno(2.12)
$$
$$
\begin{array}{l}
m_1^2 =\frac{1}{2} \left( B+C+D - \sqrt{ 8 A^2
+(B+C-D)^2 } \right) , \\
m_2^2 =\frac{1}{2} \left( B+C+D + \sqrt{ 8 A^2
+(B+C-D)^2 } \right) , \\
m_3^2= B-C .
\end{array}
\eqno(2.13)
$$
As a result, the Hermitian matrix $M M^\dagger$ is diagonalized
by
$$
U = P_{23}(\delta) P_1(\phi) R(\theta) ,
\eqno(2.14)
$$
as
$$
U^\dagger M M^\dagger U = \widetilde{H}_D .
\eqno(2.15)
$$

Since we did not considered the size of masses, the 
ordering of them is needed. Therefore, the unitary matrix to 
diagonalize the mass matrix in an proper mass ordering is 
given by $UT$, where $T$ is the matrix to exchange the mass 
ordering. Then, we find the neutrino mixing matrix 
defined by (1.7) as
$$
U = U_e^\dagger U_\nu =
T_e^T R^T(\theta_e) P_e^\dagger P_\nu R(\theta_\nu)T_\nu ,
\eqno(2.16)
$$
where
$$
P_f = P_{23}(\delta_f) P_1(\phi_f)  
={\rm diag} ( e^{i\phi_f}, e^{i \delta_f},
e^{-i \delta_f}) .
\eqno(2.17)
$$
Here, we recall that the operation (2.1) must be the same for 
$\nu_L$ and $e_L$, so that , in the expression,  $\delta_e$
is exactly equal to $\delta_\nu$.
Therefore, we obtain
$$
U = T_e^T U_0 T_\nu \equiv T_e^T \left(
\begin{array}{ccc}
s_e s_\nu +c_e c_\nu e^{i \phi} & -s_e c_\nu +c_e s_\nu e^{i \phi}  & 0 \\
-c_e s_\nu +s_e c_\nu e^{i \phi} & c_e c_\nu +s_e s_\nu e^{i \phi}  & 0 \\
0 & 0 & 1
\end{array} \right)T_\nu ,
\eqno(2.18)
$$ 
where $\phi=\phi_\nu -\phi_e$.  
Obviously, the mixing matrix (2.18) cannot give the observed values  
\cite{atm,solar} $\tan^2 \theta_{12}\simeq 1/2$ and $\sin^2 2\theta_{23} 
\simeq 1$ simultaneously. 
(It is a general feature for any flavor symmetry with a transformation
$f_L \rightarrow U_X f_L$ that we obtain only a family-mixing
between two families.  See Ref.\cite{Koide05}.)


\vspace{3mm}
\noindent{\large\bf 3 \ Extended  $2\leftrightarrow 3$ symmetry with 
the breaking term}

We saw in the previous section that the $2\leftrightarrow 3$ symmetry 
which arises  as a consequence of the transformation for fields 
cannot reproduce the observed neutrino mixing. However, 
we consider that the $2\leftrightarrow 3$ 
symmetry is still useful from the phenomenological point of view.
Therefore, from the phenomenological point
of view, we assume \cite{Koide02} that the $2\leftrightarrow 3$ 
symmetry is broken only through the phase parameters.
Hereafter, we will use the extended $2\leftrightarrow 3$ symmetry
operator (2.1) in the phenomenological meaning, and we will 
consider the case $\delta_e \neq \delta_\nu$ in the left-handed
sectors.

\vskip 3mm
\noindent
{\bf 3.1) Charged lepton mass spectrum}

First, we investigate the $2\leftrightarrow 3$ symmetry of Type II. 
The mass matrix $M_L^e$ for the charged leptons must also
satisfy the relation
$$
T_{23} (2 \delta)  M_L^e = M_L^e ,
\eqno(3.1)
$$
where, for convenience, we have dropped the index ``e" from 
$\delta_e$.
Then, the explicit form of $M_L^e$ is also given by 
$$
M_L^e = \left(
\begin{array}{ccc}
a_1 & b_1 & c_1  \\
a e^{i\delta}  &  b e^{i\delta}  &  c e^{i\delta}  \\
a e^{-i\delta}   &  b e^{-i\delta}  &  c e^{-i\delta} 
\end{array} \right) = P_{23} (\delta) \left(
\begin{array}{ccc}
a_1 & b_1 & c_1  \\
a   &  b  &  c  \\
a   &  b  &  c 
\end{array} \right) ,
\eqno(3.2)
$$
where the parameters $a,b,\cdots$ in $M_L^e$ can be complex. 
Therefore, we obtain the Hermitian matrix
$$
M_L^e (M_L^e)^\dagger = P_{23}(\delta)
\left(
\begin{array}{ccc}
D & A  e^{i\phi} & A e^{i\phi}  \\
A e^{-i\phi}   & B  &  B  \\
A e^{-i\phi}   & B  &  B 
\end{array} \right)  P_{23}^\dagger (\delta),
\eqno(3.3)
$$
where
$$
\begin{array}{l}
A= |a a_1^* +b b_1^* +c c_1^*| , \\
B=|a|^2 +|b|^2 +|c|^2 , \\
D=|a_1|^2+|b_1|^2+|c_1|^2 .
\end{array}
\eqno(3.4)
$$
Then, we can obtain a real matrix $\widetilde{H}_e$ as
$$
\widetilde{H}_e = P_1^\dagger (\phi) P_{23}^\dagger(\delta) 
M_L^e (M_L^e)^\dagger P_{23}(\delta) P_1(\phi) .
\eqno(3.5)
$$
From the formula (2.13), we obtain 
$$
m_{e3}=0 ,
\eqno(3.6)
$$
because of $B=C$ in this case.
Therefore, Type II transformation in charged lepton sector
cannot give a realistic mass spectrum.

Next, we investigate the case of Type I, i.e.
$$
\ell_L \rightarrow T_{23}(2\delta_L) \ell_L , \ \ \  
e_R \rightarrow T_{23}(2\delta_R) e_R .
\eqno(3.7)
$$ 
The case (3.5) may be realized in an SU(5)-GUT model.
In this case, instead of the constraint (3.1), we have
the constraint
$$
T_{23}(2\delta_L) M_L^e T_{23}^\dagger (2\delta_R) = M_L^e .
\eqno(3.8)
$$
The explicit form of $M_L^e$ is given by
$$
M_L^e = \left(
\begin{array}{ccc}
d & a' e^{-i\delta_R} & a' e^{i\delta_R}  \\
a e^{i\delta_L}  &  b e^{i(\delta_L-\delta_R)}  &  c e^{-i(\delta_L+\delta_R)}   \\
a e^{-i\delta_L}   &  c e^{i(\delta_L+\delta_R)}   &  b e^{-i(\delta_L-\delta_R)} 
\end{array} \right)  =
P_{23}(\delta_L)\left(
\begin{array}{ccc}
d & a' & a'  \\
a   &  b  &  c  \\
a   &  c  &  b 
\end{array} \right) P_{23}^\dagger (\delta_R) ,
\eqno(3.9)
$$
so that we obtain
$$
M_L^e (M_L^e)^\dagger = P_{23}(\delta_L)
\left(
\begin{array}{ccc}
D & A e^{i\phi} & A e^{i\phi}  \\
A e^{-i\phi}   & B  &  C  \\
A e^{-i\phi}  & C  &  B 
\end{array} \right) P_{23}^\dagger (\delta_L) ,
\eqno(3.10)
$$
where
$$
\begin{array}{l}
A= |a d^* +(b+c) a^{\prime *}| , \\
B=|a|^2 +|b|^2 +|c|^2 , \\
C=|a|^2 + 2|b||c|\cos(\beta-\gamma) , \\
D=|d|^2+ 2|a'|^2 ,
\end{array}
\eqno(3.11)
$$
where $\beta$ and $\gamma$ are defined by
$b=|b| e^{i\beta}$ and $c=|c| e^{i\gamma}$, respectively.
Therefore, since 
$$
m_{e3}^2 = B-C =|b|^2+|c|^2 
- 2 |b||c|\cos(\beta-\gamma)=|b-c|^2 ,
\eqno(3.12)
$$
we can obtain $m_{e3} \neq 0$
when $b \neq c$.

In both cases, Types I and II, the Hermitian matrix 
$M_L^e (M_L^e)^\dagger$ is diagonalized by the unitary matrix
$$
U_e = P_{23}(\delta_e) P_1(\phi_e) R(\theta_e) ,
\eqno(3.13)
$$
as
$$
U_e^\dagger M_L^e (M_L^e)^\dagger U_e = D^2_e
\equiv (m^2_{e 1},m^2_{e 2},m^2_{e 3}).
\eqno(3.14)
$$

\vspace{1mm}
\noindent
{\bf 3.2) Neutrino mass spectrum}

We consider that the neutrino masses are generated by a seesaw 
mechanism
$$
M_\nu  = M_L^\nu M_R^{-1} (M_L^\nu) ^T,
\eqno(3.15)
$$
where $M_L^\nu$ and $M_R$ are defined by $\bar{\nu}_L M_L^\nu \nu_R$
and $\bar{\nu}_R^c M_R \nu_R$ ($\nu_R^c \equiv C\bar{\nu}_R^T$), 
respectively.
The Dirac mass matrix $M_L^\nu$ is given by the form similar to (3.9) 
or (3.2) according as Type-I or Type-II.
In Type-I, we obtain the neutrino mass matrix form 
$$
M_\nu = P_{23}(\delta) \left(
\begin{array}{ccc}
D & A & A \\
A & B & B \\
A & B & B
\end{array} \right) P_{23}(\delta) ,
\eqno(3.16)
$$
where
$$
\begin{array}{l}
A=a a_1 d_R^{-1}+ b b_1 b_R^{-1} c c_1 b_R^{\prime -1} +
(a b_1+a_1 b) a_R^{-1} + (a c_1 +a_1 c) a_R^{\prime -1}
+(b_1 b c_1 +b c_1) c_R , \\
B= b^2 b_R^{-1} + c^2 b_R^{\prime -1} + a^2 d_R + 2 bc c_R^{-1}
+2 a b a_R^{-1} + 2 a c a_R^{\prime -1} \\
D= b_1^2 b_R^{-1} + c_1^2 b_R^{\prime -1} + a_1^2 d_R + 2 b_1c_1 c_R^{-1}
+2 a_1 b_1 a_R^{-1} + 2 a_1 c_2 a_R^{\prime -1} 
\end{array}
\eqno(3.17)
$$
$$ 
M_R^{-1} = \left(
\begin{array}{ccc}
d_R^{-1}  & a_R^{-1}  &  a_R^{\prime -1}  \\
a_R^{-1}  & b_R^{-1}  & c_R^{-1}   \\
a_R^{\prime -1}  & c_R^{-1}  &  b_R^{\prime -1}  
\end{array} \right) .
\eqno(3.18)
$$
Since the neutrino masses $m_{\nu i}$ in Type-II are given by
$$
\begin{array}{l}
m_{\nu 1} =\frac{1}{2} \left( B+C+D - \sqrt{ 8 A^2
+(B+C-D)^2 } \right) , \\
m_{\nu 2} =\frac{1}{2} \left( B+C+D + \sqrt{ 8 A^2
+(B+C-D)^2 } \right) , \\
m_{\nu 3}= B-C ,
\end{array}
\eqno(3.19)
$$
with $C=B$,  we obtain
$$
m_{\nu 3} = 0 .
\eqno(3.20)
$$
On the other, in Type I, such the constraint (4.6) does not appear.

In both cases, Types I and II, the Hermitian matrix 
$M_\nu M_\nu^\dagger$ is diagonalized by the unitary matrix
$$
U_\nu = P_{23}(\delta_\nu) P_1(\phi_\nu) R(\theta_\nu) ,
\eqno(3.21)
$$
as
$$
U_\nu^\dagger M_\nu M_\nu^\dagger U_\nu = D^2_\nu
\equiv (m^2_{\nu 1},m^2_{\nu 2},m^2_{\nu 3}),
\eqno(3.22)
$$
where $R(\theta_\nu)$ is defined by Eq.(2.11) with
$$
s_\nu \equiv \sin\theta_\nu =
\sqrt{\frac{D-m_{\nu 1}}{m_{\nu 2} -m_{\nu 1}}} , \ \ \ 
c_\nu \equiv \cos\theta_\nu =
\sqrt{\frac{m_{\nu 2} - D}{m_{\nu 2} -m_{\nu 1}}} . 
\eqno(3.23)
$$

\vspace{1mm}
\noindent
{\bf 3.3) Neutrino mixing matrix}

So far, we have used the notation $(f_1, f_2, f_3)$ for 
the mass eigenstates of the fundamental fermions $f$,
whose masses $m_{fi}$ have been defined by Eq.(2.13).
Hereafter, in order to distinguish the mass-eigenstates
$(e, \mu, \tau)$ and $(\nu_1, \nu_2, \nu_3)$ in the 
conventional notations from the mass-eigenstates 
whose masses $m_i$ are defined by Eq.(2.13), 
we denote the states whose masses are defined by Eq.(2.13)
as $f_i^0$.
The states $(\nu_1, \nu_2, \nu_3)$ and 
$(\nu_e, \nu_\mu, \nu_\tau)$, which is the SU(2)$_L$ partner 
of the charged lepton state $(e, \mu, \tau)$, are related by  
$$
\left(
\begin{array}{c}
\nu_e \\
\nu_\mu \\
\nu_\tau
\end{array} \right)
= U\left(
\begin{array}{c}
{\nu}_1 \\
{\nu}_2 \\
{\nu}_3
\end{array} \right) ,
\eqno(3.24)
$$
with the neutrino mixing matrix $U$ in the conventional notation.
Here, the neutrino mixing matrix $U$ in Eq.(3.24) is given by
$$
U =U_e^\dagger U_\nu .
\eqno(3.25)
$$
On the other hand, as seen in Secs.2 and 3, the mass matrices
$M_\nu M_\nu^\dagger$ and $M_L^e (M_L^e)^\dagger$ are
diagonalized by unitary matrices (3.21) and (3.13) (we denote
them $U_{0\nu}$ and $U_{0e}$), respectively.
When we define the mixing matrix
$$
U_0 = U_{0e}^\dagger U_{0\nu} =
R^T(\theta_e) P R(\theta_\nu),
\eqno(3.26)
$$
where 
$$
P = {\rm diag}(e^{i\phi}, e^{i\delta}, e^{-i\delta}) ,
\eqno(3.27)
$$
$\phi=\phi_\nu -\phi_e$ and $\delta=\delta_\nu -\delta_e$.
the mixing matrix $U_0$ does not always denote the observed
neutrino mixing matrix $U$.
When we define the observed fermions $(e, \mu, \tau)$ and 
$({\nu}_1, {\nu}_2,{\nu}_3)$ as
$$
\left(
\begin{array}{c}
{\nu}_1 \\
{\nu}_2 \\
{\nu}_3
\end{array} \right) = T_{ijk} \left(
\begin{array}{c}
{\nu}_1^0 \\
{\nu}_2^0 \\
{\nu}_3^0
\end{array} \right), \ \ \ \ 
\left(
\begin{array}{c}
e \\
\mu \\
\tau
\end{array} \right) = T_{lmn} \left(
\begin{array}{c}
e_1^0 \\
e_2^0 \\
e_3^0
\end{array} \right),
\eqno(3.28)
$$
the observed neutrino mixing matrix $U$ is given by
$$
U = T_{lmn} U_0 T_{ijk}^T ,
\eqno(3.29)
$$
where $T_{ijk}$ denotes the exchange operator 
$(f_1^0,f_2^0,f_3^0) \rightarrow (f_i^0,f_j^0,f_k^0)$.
However, as we discuss below, the possible choices
of $T_{ijk}$ are not so many.

The explicit form of the matrix $U_0$ is given by
$$
U_0 = \left(
\begin{array}{ccc}
c_e c_\nu e^{i\phi} + s_e s_\nu \cos\delta  & 
c_e s_\nu e^{i\phi} - s_e c_\nu \cos\delta &
is_e \sin\delta \\
s_e c_\nu e^{i\phi} - c_e s_\nu \cos\delta & 
s_e s_\nu e^{i\phi} + c_e c_\nu \cos\delta &
-ic_e \sin\delta \\
i s_\nu \sin\delta & -i c_\nu \sin\delta & \cos\delta
\end{array} \right) .
\eqno(3.30)
$$
Obviously, the cases (3.29) with $\delta=0$ are ruled out
as we have already discussed in Sec.2.


\begin{table}
\caption{
Possible constraints on the Dirac mass matrices $m_L^f$:
Models A, B, C, and D are defined according as the constraint types.
}

\begin{center}
\begin{tabular}{|c|l|l|}\hline
Type  & \ \ Type II for $M_L^\nu$  & \ Type I for $M_L^\nu$ \\ \hline
Type II for $M_L^e$  &  Model A: non-GUT type  & 
Model D: unrealistic  \\ 
                     &  $m_{e3}=m_{\nu 3}=0$  &
$m_{e3}= 0$ \& $m_{\nu 3}\neq 0$ \\ \hline
Type I for $M_L^e$  &  Model B: SU(5)-GUT type &
Model C: SO(10)-GUT type \\
          &  $m_{e3}\neq 0$ \& $m_{\nu 3}= 0$ &
$m_{e3}\neq 0$ \& $m_{\nu 3}\neq  0$ \\ \hline
\end{tabular}
\end{center} 
\end{table}


For convenient, we name Models A, B, C and D for combinations of Types I and II
for $M_L^e$ and $M_L^\nu$ as shown in Table 1.
In Model A, since only the left-handed fields $f_L$ obey the
$2\leftrightarrow 3$ symmetry, the model cannot be embedded into a
GUT scenario.
In Model B, the fields $\ell_L =(\nu_L, e_L)$ and $e_R$ obey the
$2\leftrightarrow 3$ symmetry, but the field $\nu_R$ is free from
the symmetry, so that the model can be embedded into SU(5) GUT.
In Model C, all fields $\ell_L =(\nu_L, e_L)$, $e_R$ and $\nu_R$ 
obey the $2\leftrightarrow 3$ symmetry, so that the model can be
embedded into SO(10) GUT.
Model D is unlikely, so that we will not investigate this case.

 In Models A and D with Type-II symmetry in the charged lepton sector,
 we obtain $m_{e3}=0$, so that the cases are ruled out.
 
In  Model B (a SU(5)-GUT-type model), we can obtain $m_{e3}\simeq 0$ 
(but $m_{e3}\neq 0$) because of $b\simeq c$.
(In Model B, although we can, in principle, consider any value 
of $m_{e3}$, we have assumed $b\simeq c$ because
the case $b\simeq c$ can reasonably be realized 
in most practical models with 
$2\leftrightarrow 3$ symmetry.)     
Therefore, we may suppose a case $m_{e2}^2 > m_{e1}^2 > m_{e3}^2$
in the model. 
Such the case means the assignment
$$
\left(
\begin{array}{c}
e \\
\mu \\
\tau
\end{array} \right)_L = 
\left(
\begin{array}{c}
e_3^0 \\
e_1^0 \\
e_2^0
\end{array} \right)_L  =
\left(
\begin{array}{ccc}
0 & 0 & 1 \\
1 & 0 & 0 \\
0 & 1 & 0
\end{array} \right) \left(
\begin{array}{c}
e_1^0 \\
e_2^0 \\
e_3^0
\end{array} \right)_L \equiv T_{312} \left(
\begin{array}{c}
e_1^0 \\
e_2^0 \\
e_3^0
\end{array} \right)_L
.
\eqno(3.31)
$$
Then, from the relation $U_e = U_{e0} T_{312}^T$, 
the observed neutrino mixing matrix $U$ is described 
by
$$
U=T_{312} U_0 = \left(
\begin{array}{ccc}
i s_\nu \sin\delta & -i c_\nu \sin\delta & \cos\delta \\
c_e c_\nu e^{i\phi} + s_e s_\nu \cos\delta  & 
c_e s_\nu e^{i\phi} - s_e c_\nu \cos\delta &
i s_e \sin\delta \\
s_e c_\nu e^{i\phi} - c_e s_\nu \cos\delta & 
s_e s_\nu e^{i\phi} + c_e c_\nu \cos\delta &
-ic_e \sin\delta 
\end{array} \right) .
\eqno(3.32)
$$
if we regard the observed neutrino states   
$({\nu}_1, {\nu}_2,{\nu}_3)$ as $(\nu_1^0, \nu_2^0, \nu_3^0)$
with $m_{\nu_3}=0$, whose case corresponds to
the inverse hierarchy.
(Such an inverted assignment between up- and down-sectors was 
first proposed by Matsuda and Nishiura \cite{Matsuda04}.)
The case (3.32) predicts
$$
\tan^2 \theta_{solar} = \frac{|U_{12}|^2 }{|U_{11}|^2}
=\frac{c_\nu^2 }{s_\nu^2}
= \frac{m_{\nu 2} -D_\nu}{D_\nu -m_{\nu 1}},
\eqno(3.33)
$$
$$
\sin^2 2\theta_{atm} = 4|U_{23}|^2 |U_{33}|^2 =
 \sin^2 2\theta_e \sin^4\delta 
= \sin^2 2\theta_e (1-|U_{13}|^2)^2,
\eqno(3.34)
$$
where $s_e$ and $c_e$ are given by Eq.(2.12).
In order to give $|U_{13}|^2 \simeq 0$,  the condition 
$\cos\delta \simeq 0$  ($\delta\simeq \pi/2$) is required.
In order to $\sin^2 2\theta_e=1$ ($s_e^2=c_e^2=1/2$),  the relation
$2D_e=m_{e1}^2+m_{e2}^2$ (i.e. $D_e=B_e+C_e$) is required
from Eq.(2.13).
Then, the masses (2.13) are given by
$$
\begin{array}{l}
m_{e3}^2=B_e-C_e=|b_e-c_e|^2 , \\
m_{e1}^2 = D_e - \sqrt{2} A_e , \\
m_{e2}^2= D_e+\sqrt{2} A_e .
\end{array}
\eqno(3.35)
$$
Therefore, a suitable choice of the parameter values of $M_L^e$ can give 
$\sin^2 2\theta_e=1$ keeping $m_{e2}^2> m_{e1}^2>m_{e3}^2$.
Also, a suitable choice of the parameter values of $M_\nu$ can give
a reasonable value of (3.33).
If these conditions are satisfied, the model B is preferable.
However, note that the parameter value $\delta \simeq \pi/2$ cannot be
realized unless SU(2)$_L$ is broken.

By the way, the case $m_{\nu 3}=0$ does not always mean the inverse
hierarchy of neutrino masses.
At present, as far as the observed neutrino masses ${m}_{\nu_i}$ 
satisfy the relation 
$({m}_{\nu 2}^2-{m}_{\nu 1}^2)/|({m}_{\nu 3}^2-{m}_{\nu 2}^2)| \sim 10^{-2}$,
we may consider any cases $U=T_{312} U_0 T_{ijk}^T$.
Therefore, even the case $m_{\nu 3}=0$, we can consider a case of the normal 
hierarchy: $({\nu}_1, {\nu}_2, {\nu}_3)=(\nu_3^0, \nu_1^0, \nu_2^0)$.
Then, in  Model B with $c_e\simeq b_e$, 
the neutrino mixing matrix $U$ is given by  
$$
U = T_{312} U_0 T_{312}^T
= \left(
\begin{array}{ccc}
 \cos\delta & i s_\nu \sin\delta & -i c_\nu \sin\delta  \\
 i s_e \sin\delta & 
c_e c_\nu e^{i\phi} + s_e s_\nu \cos\delta  & 
c_e s_\nu e^{i\phi} - s_e c_\nu \cos\delta  \\
-ic_e \sin\delta & 
s_e c_\nu e^{i\phi} - c_e s_\nu \cos\delta & 
s_e s_\nu e^{i\phi} + c_e c_\nu \cos\delta 
\end{array} \right) .
\eqno(3.36)
$$
In order to give $\tan^2 \theta_{solar} \simeq 1/2$ and
$\sin^2 2\theta_{atm} \simeq 1$, we have to consider
$c_\nu \simeq 0$.
From the expression (3.23), the limit of $c_\nu = 0$
requires $m_{\nu 2}=D_\nu$, which leads $A_\nu=0$ 
and gives the mass spectrum $m_{\nu 1} =D_\nu$, 
$m_{\nu 2}=2B_\nu$ and $m_{\nu 3}=0$.
If we choose $B_\nu^2 \gg D_\nu^2$ in the neutrino sector, 
we can give a reasonable value of 
$R=\Delta m^2_{solar}/\Delta m^2_{atm}$
because of $R=(m_1^2-m_3^2)/(m_2^2-m_1^2)=
D_\nu^2/(4B_\nu^2-D_\nu^2)$
in the normal mass hierarchy.
Therefore, we cannot rule out this case (Model B with
$m_{e 2}^2 \gg m_{e 1}^2 \gg m_{e 3}^2$ and
$m_{\nu 2}^2 \gg m_{\nu 1}^2 \gg m_{\nu 3}^2$
in a normal hierarchy).
However, we must accept a phenomenological value
$\tan^2\delta \simeq 1/2$ ($\delta \simeq 35.3^\circ$)
in order to understand $\tan^2\theta_{solar}\simeq 1/2$.

So far, we have consider the case with $c_e\simeq b_e$
(i.e. $m_{e3}^2 \ll m_{e1}^2 \ll m_{e2}^2$) for the charged 
lepton masses in Model B.
We can also consider the case 
$m_{e1}^2 \ll m_{e2}^2 \ll m_{e3}^2$ in Model B.
In Model B, the neutrino masses are still given by
$m_{\nu 3}^2=0 < m_{\nu 1}^2 < m_{\nu 2}^2$, so that 
the cases $U=T_{123}U_0 T_{312}^T$ and $U=T_{123}U_0 T_{123}^T$
correspond to the normal and inverse hierarchies, respectively.
The explicit form of $U$ for the case $U=T_{123}U_0 T_{123}^T$ 
has been given in (3.30) because $U=T_{123}U_0 T_{123}^T=U_0$.
The explicit form of the case $U=T_{123}U_0 T_{312}^T$ is given by
 $$
U_0 = \left(
\begin{array}{ccc}
c_e c_\nu e^{i\phi} + s_e s_\nu \cos\delta  & 
c_e s_\nu e^{i\phi} - s_e c_\nu \cos\delta &
is_e \sin\delta \\
s_e c_\nu e^{i\phi} - c_e s_\nu \cos\delta & 
s_e s_\nu e^{i\phi} + c_e c_\nu \cos\delta &
-ic_e \sin\delta \\
i s_\nu \sin\delta & -i c_\nu \sin\delta & \cos\delta
\end{array} \right) .
\eqno(3.37)
$$
In order to see whether those cases cannot be ruled out or
not, it is convenient to see whether we can take or not 
possible parameter values in the limit of 
$\tan^2 \theta_{solar}=1/2$, $\sin^2 2\theta_{atm}=1$ 
and $|U_{13}|^2=0$, without contradicting with the
observed neutrino mass hierarchy.
The results are listed in Table 2.
All cases are acceptable if we neglect the problem 
whether such a set of the parameter values is 
natural or not, although we think that the case with
$U=T_{123}U_0 T_{312}^T$ is unlikely.

\begin{table}
\caption{
Possible neutrino mixing matrix form in Model B.
}

\begin{center}
\begin{tabular}{|c|cc|cc|}\hline
$m_{\nu 0i}$ & \multicolumn{4}{c|}{$m_{\nu 03}^2=0
< m_{\nu 01}^2 < m_{\nu 02}^2$} \\ \hline
$m_{e 0i}$ & \multicolumn{2}{c|}{$m_{e 03}^2
< m_{e 01}^2 < m_{e 02}^2$} & 
\multicolumn{2}{c|}{$m_{e 01}^2 < m_{e 02}^2 < m_{e 03}^2$} 
\\ \hline
Hierarchy & Normal &  Inverse &  Normal  & Inverse \\
$U$ & $T_{312}U_0 T_{312}^T$ & $T_{312}U_0 T_{123}^T$ &
$T_{123}U_0 T_{312}^T$ &  $T_{123}U_0 T_{123}^T$ \\
\hline
Limit of & $\tan^2 \delta=1/2$ & $\delta=\pi/2$ &
$\tan^2\delta=5$ & $\delta=\pi/4$ \\
$\sin^2 2\theta_{23}=1$  & $s_e^2=1/2$ & $s_e^2=1/2$ & $s_e^2 =4/5$ & 
$s_e^2=0$ \\
\& $\tan^2\theta_{12}=1/2$ & $s_\nu^2=1$ & $s_\nu^2 =2/3$ & 
$s_\nu^2 =2/5$ & $s_\nu^2 =1/3$ \\ \hline
\end{tabular}
\end{center} 
\end{table}

In Model C, since we can take any order of $m_i^2$, we cannot
say any definite conclusion (predictions) without giving
the explicit mass matrix parameters.
Therefore, for the case C, we do not give a table  such as 
Table 2.

\vspace{3mm}
\noindent{\large\bf 4 \ Summary}

In conclusion, we have systematically investigated possible
lepton mass mass matrix forms and mixings under the
expended $2\leftrightarrow 3$ symmetry.
We gave investigated two types of the $2\leftrightarrow 3$ symmetry:
one (Type I) is  that the left- and right-handed fields $(f_L, f_R)$ 
obey the symmetry, and another one (Type II) is that  only $f_L$ 
obeys the symmetry.
Note that even in Type II, in spite of no $2\leftrightarrow 3$ 
symmetry  in the
Majorana mass matrix $M_R$ for $\nu_R$, the neutrino seesaw mass
matrix still obey the $2\leftrightarrow 3$ symmetry.  
However, we have concluded that the fermion mass $m_3$ is always zero 
in Type II.
Therefore, the possibility that the charged lepton sector obeys the 
$2\leftrightarrow 3$ symmetry of Type II is ruled out.
We have been interested in the case B classified in Table 1, where
the neutrino sector obeys  the $2\leftrightarrow 3$ symmetry of Type II,
because we consider a model with an SU(5)-GUT type scenario 
\cite{Mohapatra06}.
In this case, we have only four cases of the neutrino mixing matrix.
The results are summarized in Table 2.

We are also interested in a model with an SO(10)-type scenario.
In this case (Model C),  the right-handed neutrino $\nu_{R}$ is
also  transformed as $\nu_R \rightarrow T_{23} \nu_R$,
so that we can consider any value of $m_{\nu_{03}} \neq 0$
and any mixing matrix form (2.19).
However, in the SO(10)-GUT model, a more strict constraint on
the neutrino mass matrix appears because the neutrino mass matrix
form is strictly related to the quark and charged lepton mass matrices, 
so that most naive SO(10) models have, at present, not succeeded 
\cite{SO10} in giving reasonable fits for all the masses and 
mixings in the quark and lepton sectors, even without
the $2\leftrightarrow 3$ symmetry.

In the practical point of view, we think that there is a possibility
to build a realistic model  based on SU(5)-GUT rather than
SO(10).
In Model B, we are interested in the case of an inverse neutrino 
mass hierarchy, because the case $\delta =\pi/2$ is likely.
The case predicts the effective electron neutrino mass 
$\langle m_{\nu e}\rangle$ is of the order of 
$\sqrt{\Delta m^2_{atm}} \simeq 0.05$ eV, which is
within the reach of the next generation experiments of the 
neutrinoless double beta decay.

We hope that the present investigation will be helpful
to investigate more explicit model based on a GUT scenario.

\vspace{4mm}

\centerline{\large\bf Acknowledgment} 

The one of the authors (YK) is supported by the Grant-in-Aid for
Scientific Research, Ministry of Education, Science and 
Culture, Japan (No.18540284).


\end{document}